# Restricted Conformal Property of Compressive Sensing

Tao Cheng

*[The work of T. Cheng is supported by the National Natural Science Foundation of China under Grants 41461082.*
*T. Cheng is with the Department of Mechanics, Automotive & Transportation Engineering Institute, Guangxi University of Science and Technology, Liuzhou, Guangxi 545006 P. R. China (e-mail: ctnp@163.com).*

*Corresponding author: Tao Cheng, Tel: +86-772-268-6163, E-Mail: ctnp@163.com]*

***Abstract*—Energy and direction are tow basic properties of a vector. A discrete signal is a vector in nature. RIP of compressive sensing can not show the direction information of a signal but show the energy information of a signal. Hence, RIP is not complete. Orthogonal matrices can preserve angles and lengths. Preservation of length can show energies of signals like RIP do; and preservation of angle can show directions of signals. Therefore, Restricted Conformal Property (RCP) is proposed according to preservation of angle. RCP can show the direction of a signal just as RIP shows the energy of a signal. RCP is important supplement and development of RIP. Tow different proofs of RCP are given, namely, RCP_JL and RCP_IP.**

## I. INTRODUCTION

### A. *Incomplete of RIP*

A discrete signal is a vector in nature. Energy and direction are tow basic properties of a vector. An orthogonal matrix $\mathbf{A}$ can preserve angles ( $\frac{\langle \mathbf{Au}, \mathbf{Av} \rangle}{\|\mathbf{Au}\| \cdot \|\mathbf{Av}\|} = \frac{\langle \mathbf{u}, \mathbf{v} \rangle}{\|\mathbf{u}\| \cdot \|\mathbf{v}\|}$ ) and lengths or energies ( $\|\mathbf{Au}\| = \|\mathbf{u}\|$ or $\|\mathbf{Au}\|^2 = \|\mathbf{u}\|^2$ ) where $\mathbf{u}, \mathbf{v} \in R^N$ and $\mathbf{A} \in R^{N \times N}$. Although directions of $\mathbf{Au}$ and $\mathbf{u}$ are not same, the included angle between $\mathbf{Au}$ and $\mathbf{Av}$ is equal to the included angle between $\mathbf{u}$ and $\mathbf{v}$. Compared to $\mathbf{u}$ and $\mathbf{v}$, relative directions between $\mathbf{Au}$ and $\mathbf{Av}$ are unchanged. Hence, $\mathbf{Au}$ and $\mathbf{Av}$ have inherited energies and directions of $\mathbf{u}$ and $\mathbf{v}$ simultaneously.

Measurement and reconstruction matrices of compressive sensing are ill-conditioned matrices of non-full column rank. Some design and optimization rules of measurement and reconstruction matrices were proposed. Three conditions based on independence of all small groups of columns were proposed by Donoho[1-3]. Restricted isometry property (RIP) based on restricted isometry constants (RIC) and restricted orthogonality constants (ROC) was proposed by Candes and Tao[4]. (See definition 1.1 for an illustration). Aims of all these rules are to measure how well a group of column vectors of a measurement or reconstruction matrix behaves like an orthogonal matrix. If a matrix is similar with an orthogonal matrix, it should approximately preserve angles and lengths (energies) of vectors. RIP can't show directions of vectors, although energies of vectors can be showed by RIC and the size of the principal angle between subspaces of different dimension can be controlled by ROC and RIC [4]. Hence, RIP is not complete.

Definition 1.1 (RIP)[4, 5]: Let $\mathbf{\Phi} \in \mathbf{R}^{M \times N}$ , $M < N$ , $\mathbf{x}, \mathbf{x}' \in \mathbf{R}^N$ and $I, I' \subset \{1, \cdots, N\}$ . For every integer $K, K' \in [1, N]$, we define the K-restricted isometry constants (RIC) $\delta_K$ to be the smallest quantity such that $\mathbf{\Phi}_I$ obeys

$$(1-\sigma_K)\|\mathbf{x}\|^2 \le \lambda_{\min}(\mathbf{\Phi}_I^T\mathbf{\Phi}_I)\|\mathbf{x}\|^2 \le \|\mathbf{\Phi x}\|^2 \le \lambda_{\max}(\mathbf{\Phi}_I^T\mathbf{\Phi}_I)\|\mathbf{x}\|^2 \le (1+\sigma_K)\|\mathbf{x}\|^2 \tag{1.1}$$

where $\sigma_K \in [0,1)$ . $\lambda_{\min}(\mathbf{\Phi}_I^T\mathbf{\Phi}_I)$ and $\lambda_{\max}(\mathbf{\Phi}_I^T\mathbf{\Phi}_I)$ denote the minimal and maximal eigenvalues of $\mathbf{\Phi}_I^T\mathbf{\Phi}_I$ , respectively[4, 5]. Similarly, we define the $K, K'$ -restricted orthogonality constants (ROC) $\theta_{K,K'}$ for $K + K' \le N$ to be the smallest quantity such that

$$\left|\langle \mathbf{\Phi x}, \mathbf{\Phi x}' \rangle\right| \le \theta_{K,K'} \|\mathbf{x}\|\|\mathbf{x}'\| \tag{1.2}$$

holds for all disjoint sets $|I| + |I'| \le N$ of cardinality $|I| \le K$ and $|I'| \le K'$ .

### B. *RCP Based on Linear Array Push-broom Mode*

Push broom imaging by a linear array of detectors is main data collection technology in remote sensing. When data of remote sensing imagery $\mathbf{X}$ are collected column by column by a Gaussian measurement matrix ( $\mathbf{y}_j = \mathbf{\Phi x}_j$ where $\mathbf{x}_j \in R^N, \mathbf{X} \in R^{N \times L}$ , $\mathbf{Y} \in R^{M \times L}, 0 \le j \le L$ , $\mathbf{x}_j$ is a column of $\mathbf{X}$ ), I found that measurements retain not only energies but also directions of all columns of remote sensing imagery approximatively. (See fig. 1 for an illustration, fig. 1 for sparse signals i.e. sparse images[6]. $\mu$ is the correlation coefficient between 2 columns, i.e. the included angle cosine of 2 columns. Fig. 1(c) is the image of flooded area around Mulargia Lake of Sardinia Island in Italy in July 1996. Grey values of black zones are zero in fig. 1(c).) Fig. 1(a) shows good RIP since the energy curve of $\mathbf{y}_j$ consists with $\mathbf{x}_j$. Moreover, the $\mu$ curve of $\mathbf{y}_j$ consists with $\mathbf{x}_j$ in fig. 1(b), too. Hence a new theory to depict the fitting of direction curves in fig. 1(b) is necessary.

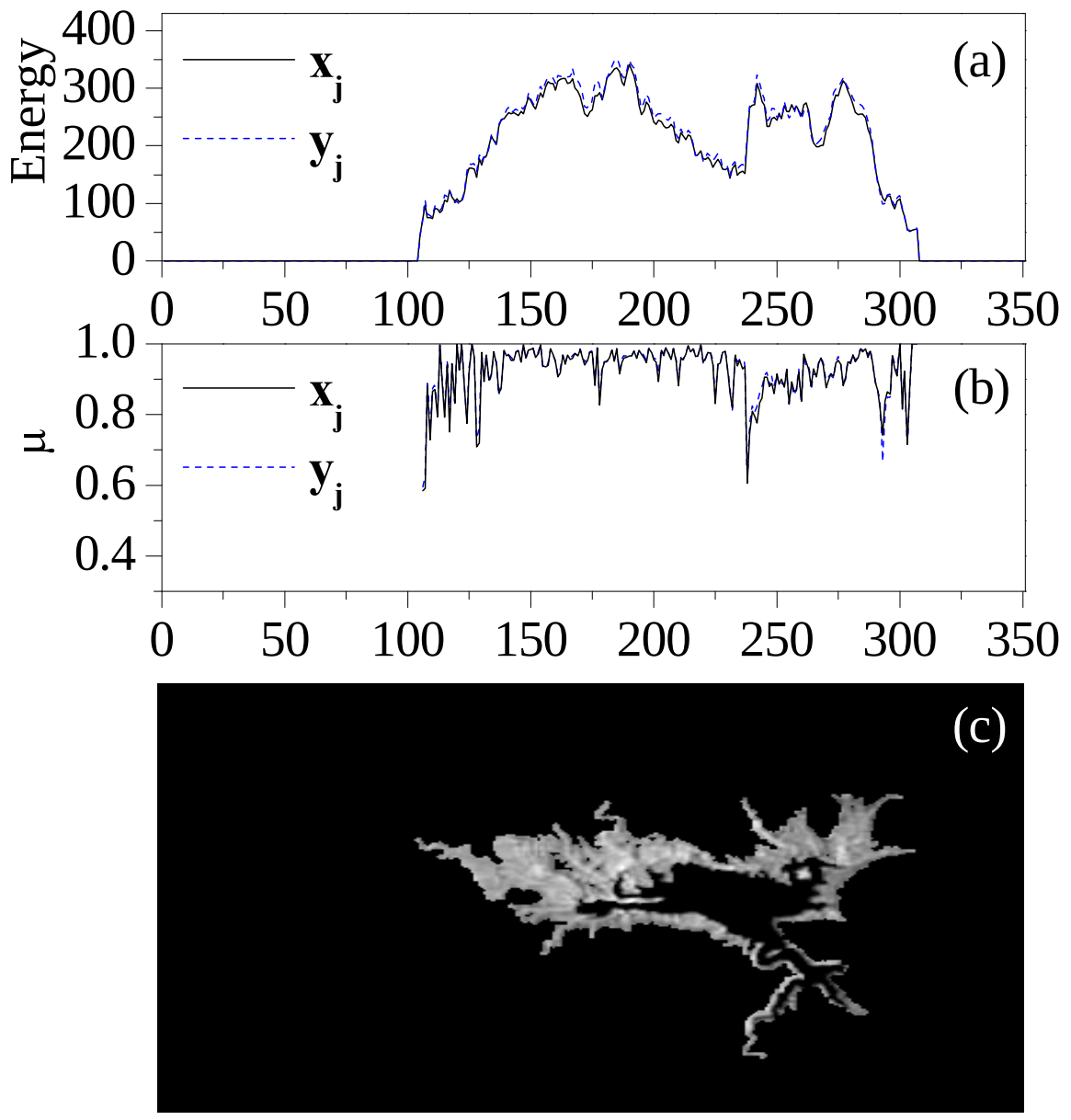

Fig. 1. Analysis curves of **X** and **Y** and gray image. (a) Energy; (b) Correlation coefficient between 2 columns; (c) Difference image generated by different temporal.

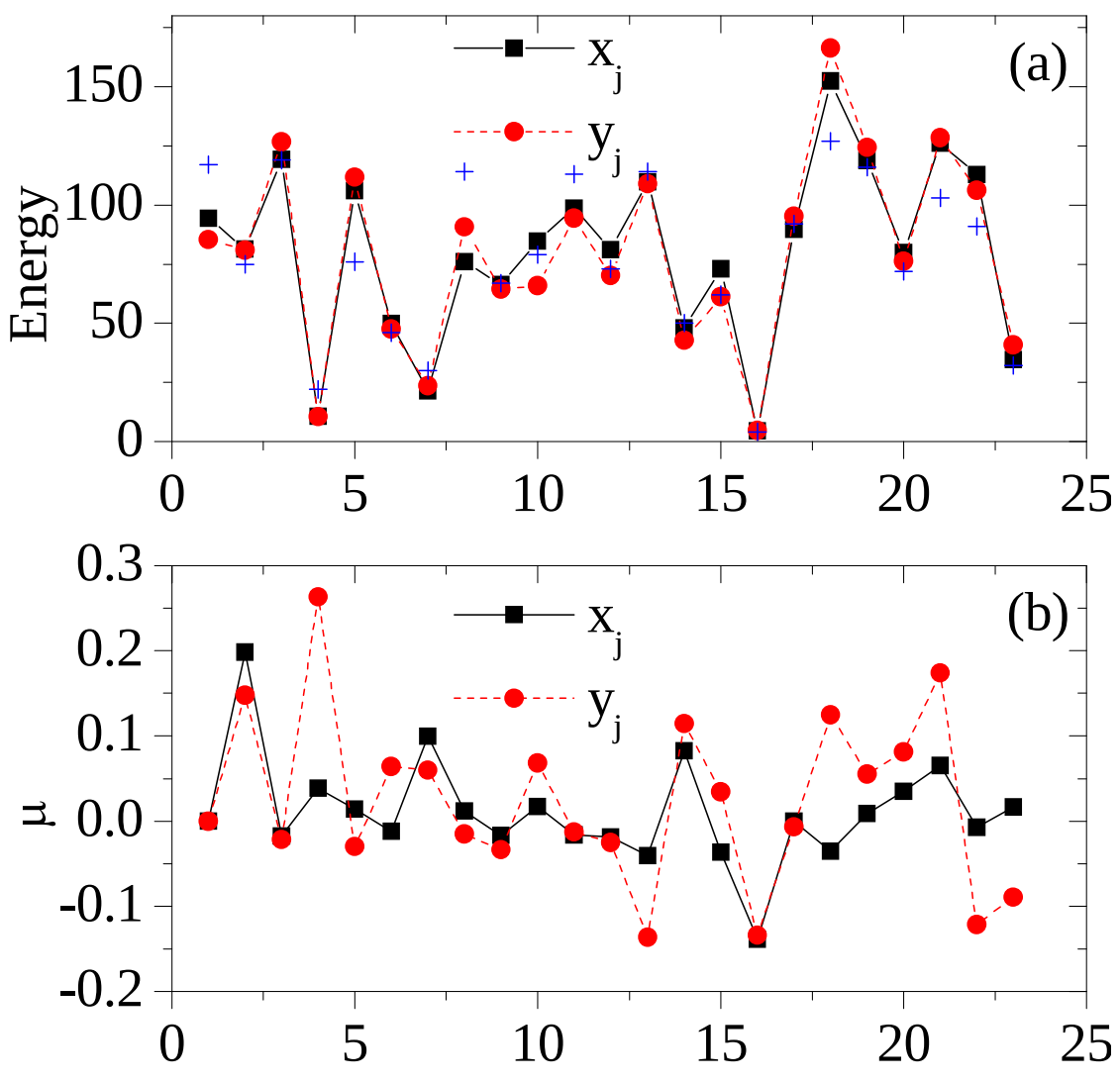

Fig. 2. Analysis curves of 23 Gaussian sparse signals and their measurements. (a) Energy; (b) Correlation coefficient between 2 Gaussian sparse signals whose sparse between 4 and 119.

Restricted conformal property (RCP) is proposed to depict the inheritance of directions in fig. 1(b) as RIP do. RCP_IP and RCP_JL are proved based on inner product and Johnson- Lindenstrauss lemma respectively. RCP_IP and RCP_JL are all called RCP.

The following definition is used in subsequent derivations.

Definition 1.2: Let $\mathbf{x}_u, \mathbf{x}_v \in \mathbf{R}^N$ be discrete sparse signals supported on unknown set $I_u$ and $I_v$ where $\xi = \frac{\|\mathbf{x}_u\|^2 + \|\mathbf{x}_v\|^2}{2\|\mathbf{x}_u\|\|\mathbf{x}_v\|} \ge 1$. Put $Ku \ge |I_u|$, $Kv \ge |I_v|$, $I = I_u \cup I_v$ and $K \ge |I_u \cup I_v|$. Let $\mathbf{\Phi}_I$ be the submatrix with column indices $j \in I$ where $\mathbf{\Phi} \in \mathbf{R}^{M \times N}$, $\delta_{\max} = \max(\delta_{Ku}, \delta_{Kv})$, $\delta_{\max} \le \delta_K$ and $\sigma_K, \sigma_{Ku}, \sigma_{Kv} \in [0,1)$. Write $\mathbf{y}_u = \mathbf{\Phi}\mathbf{x}_u$ and $\mathbf{y}_v = \mathbf{\Phi}\mathbf{x}_v$. $\cos\beta$ is the included angle cosine of $\mathbf{\Phi}\mathbf{x}_u$ and $\mathbf{\Phi}\mathbf{x}_v$, $\cos\alpha$ is the included angle cosine of $\mathbf{x}_u$ and $\mathbf{x}_v$. $\lambda_{\min}$ and $\lambda_{\max}$ denote the minimal and maximal eigenvalues of $\mathbf{\Phi}_I^T\mathbf{\Phi}_I$, respectively[4, 5].

### C. *Organization of the Paper*

The paper is organized as follows. Section Ⅱ proves RCP by Johnson-Lindenstrauss lemma. Section Ⅲ proves RCP by Inner product. Theorem 3.3 (Identical Orthant Theorem) and Theorem 3.4 (Minus Term Theorem) are proposed to estimate the conditions in which RCP_IP holds. Section Ⅳ further expands the scope of application of RCP in compressible signals and shows the potential applications of RCP in reconstruction algorithms. Section Ⅴ discusses statistical analysis of Theorem 3.4 (Minus Term Theorem) based on Gauss matrices. In section Ⅵ, the main results of the paper are summarized.

## II. PROOF OF RCP BASED ON JOHNSON-LINDENSTRAUSS LEMMA

Lemma 2.1 (Johnson-Lindenstrauss Lemma) [4, 7]: Let $\varepsilon \in (0,1)$. For any set $Q$ of $\#(Q)$ points in $\mathbf{R}^N$, if $M$ is a positive integer such that $M > O(\ln(\#(Q))/\varepsilon^2)$, there exists a Lipschitz mapping $f: \mathbf{R}^N \to \mathbf{R}^M$ such that

$$(1-\varepsilon)\|\mathbf{u}-\mathbf{v}\|^2 \le \|f(\mathbf{u}) - f(\mathbf{v})\|^2 \le (1+\varepsilon)\|\mathbf{u}-\mathbf{v}\|^2 \tag{2.1}$$

for all $\mathbf{u}, \mathbf{v} \in Q \subset \mathbf{R}^N$ .

When $\mathbf{u}$ and $\mathbf{v}$ are sparse and Lipschitz mapping $f$ in formula (2.1) is a matrix $\mathbf{\Phi} \in \mathbf{R}^{M\times N}$ , formula (2.2) holds.

$$(1-\varepsilon)\left\|\mathbf{u}-\mathbf{v}\right\|^2 \le \left\|\mathbf{\Phi u}-\mathbf{\Phi v}\right\|^2 \le (1+\varepsilon)\left\|\mathbf{u}-\mathbf{v}\right\|^2 \tag{2.2}$$

Both $\mathbf{\Phi u}-\mathbf{\Phi v}$ and $\mathbf{u}-\mathbf{v}$ in formula (2.2) are differences of vectors. Polynomials with included angle cosines of $\mathbf{\Phi u}-\mathbf{\Phi v}$ and $\mathbf{u}-\mathbf{v}$ can be gained by the polynomial expansion based on the law of cosines. Random matrices satisfy RIP, if it satisfy Johnson-Lindenstrauss lemma[7]. Hence RCP_JL is proved based on Johnson-Lindenstrauss lemma and RIP.

Theorem 2.1 (RCP_JL): If a random matrix $\mathbf{\Phi} \in \mathbf{R}^{M\times N}$ satisfies Johnson-Lindenstrauss lemma and RIP simultaneously,

$$\frac{\delta_{\max}-\varepsilon}{1-\delta_{\max}}\xi+\frac{1-\varepsilon}{1-\delta_{\max}}\cos\alpha \ge \cos\beta \ge \frac{\varepsilon-\delta_{\max}}{1+\delta_{\max}}\xi+\frac{1+\varepsilon}{1+\delta_{\max}}\cos\alpha \tag{2.3}$$

The proof of Theorem 2.1 can be found in Appendix A.

$\delta_{\max}$ and $\varepsilon$ are constants, if a measurement matrix is determined. I expect $\cos\beta$ is close to $\cos\alpha$ . $\xi$ is the main error source which leads to the difference between $\cos\beta$ and $\cos\alpha$ in formula (2.3). There are tow methods to reduce the error. First, let $\xi$ be about 1 which is the minimum value of $\xi$. Thus $\frac{\delta_{\max}-\varepsilon}{1-\delta_{\max}}\xi$ and $\frac{\varepsilon-\delta_{\max}}{1+\delta_{\max}}\xi$ play a small role in formula (2.3). Second, let $\cos\alpha$ be about 1 which is the maximum value of $\cos\alpha$ . Thus $\frac{1-\varepsilon}{1-\delta_{\max}}\cos\alpha$ and $\frac{1+\varepsilon}{1+\delta_{\max}}\cos\alpha$ play a leading role in formula (2.3) such that $\frac{\delta_{\max}-\varepsilon}{1-\delta_{\max}}\xi$ and $\frac{\varepsilon-\delta_{\max}}{1+\delta_{\max}}\xi$ play a small role in formula (2.3) comparatively. Therefore if both $\xi$ and $\cos\alpha$ are about 1, $\cos\beta$ is close to $\cos\alpha$ .

Spatio-temporal continuity is important structure information in a natural image. Adjacent columns of an image are similar, although grey values change considerably in different zones. Energies of adjacent column vectors are close such that $\xi$ is about 1 (See fig. 1(a) for more illustration.) and their included angle cosines ( $\cos\alpha$ ) are close to 1 (See fig. 1(b) for more illustration.). Hence curves of $\mathbf{x}_j$ (i.e. $\cos\alpha$ ) and $\mathbf{y}_j$ (i.e. $\cos\beta$ ) are similar in fig. 1(b).

Fig. 2 shows curves of energy and $\mu$ of 23 Gaussian sparse signals and their measurements by a Gaussian measurement matrix. “+” shows the sparse of every Gaussian sparse signal in fig. 2(a). Curves of $\mu$ are not similar in fig. 2, since $\xi$ and $\cos\alpha$ are not close to 1.

Although formula (2.3) is suitable for any vector pair of any included angle, $\mathbf{\Phi}$ is unknown and undetermined in Johnson-Lindenstrauss lemma. The measurement matrix is known in CS. Therefore I need to estimate how well $\cos\alpha$ fits $\cos\beta$ if the measurement matrix is known and determined. RCP_JL can’t tell us more. Therefore RCP_IP is derived.

## III. PROOF OF RCP BASED ON INNER PRODUCT

### *A. RCP_IP Theorem*

Theorem 3.1 (RCP_IP): (1) $\mathbf{x}_u$ and $\mathbf{x}_v$ are sparse and $\mathbf{x}_u, \mathbf{x}_v \in \mathbf{R}^N$ , $\mathbf{\Phi} \in \mathbf{R}^{M\times N}$ . If $\lambda_{\min} < \mathbf{x}_u, \mathbf{x}_v > \le < \mathbf{\Phi x}_u, \mathbf{\Phi x}_v > \le \lambda_{\max} < \mathbf{x}_u, \mathbf{x}_v >$ and $\mathbf{\Phi}$ satisfy RIP, formula (3.1) holds; (2) If $\mathbf{x}_u$ and $\mathbf{x}_v$ are sparse and orthogonal, and $\mathbf{\Phi}$ satisfy RIP, formula (3.2) holds.

$$\frac{1-\delta_K}{1+\delta_K}\cos\alpha \le \cos\beta \le \frac{1+\delta_K}{1-\delta_K}\cos\alpha \tag{3.1}$$

$$\frac{-\delta_K}{1+\delta_K} \le \cos\beta \le \frac{\delta_K}{1-\delta_K} . \tag{3.2}$$

The proof of Theorem 3.1 can be found in Appendix B[4, 5].

$\delta_K$ is the only crucial factor which controls RCP_IP as RIP does. Hence RCP_IP and RIP are similar in nature. RCP_IP is

good, if RIP is good. There is the same constant ($\delta_K$) in RCP_IP and RIP, while there are three constants and a variable, namely, $\delta_{\max}$, $\varepsilon$ and $\xi$ besides $\cos\alpha$ and $\cos\beta$ in RCP_JL. Hence RCP_IP is simpler and easier to use than RCP_JL by comparison. Adjacent column vectors are similar in natural image. Their $I_u$ and $I_v$ are similar such that $\delta_{Ku}$ and $\delta_{Kv}$ are close. Moreover $\delta_K$ and $\delta_{\max}$ are very approximate, although $\delta_K \geq \delta_{\max}$.

When $\lambda_{\min} < \mathbf{x}_u, \mathbf{x}_v > \leq < \mathbf{\Phi x}_u, \mathbf{\Phi x}_v > \leq \lambda_{\max} < \mathbf{x}_u, \mathbf{x}_v >$ holds in Theorem 3.1 (1) is the next problem to answer.

*B. Preliminaries*

Define the Gram matrix $\mathbf{G} := \mathbf{\Phi}_I^T \mathbf{\Phi}_I$. Since $\mathbf{G}$ is a symmetric matrix, it is orthogonally diagonalizable. Let $\mathbf{G}=\mathbf{V\Lambda V}^T$ where $\mathbf{V}$ is an orthogonal matrix, and $\mathbf{V}^T\mathbf{V}=\mathbf{I}$. $\left\{\lambda_1,\ldots,\lambda_{|I|}\right\}$ denotes eigenvalues of $\mathbf{\Phi}_I^T\mathbf{\Phi}_I$, $\mathbf{\Lambda} = diag(\lambda_1,\ldots,\lambda_{|I|})$, $\lambda_1 \geq \cdots \geq \lambda_{|I|}$ where $\lambda_{\max}=\lambda_1$ and $\lambda_{\min}=\lambda_{|I|}$.

Theorem 3.2: Let $\mathbf{z}_u = \mathbf{V}^T\mathbf{x}_{uI}$, $\mathbf{z}_v = \mathbf{V}^T\mathbf{x}_{vI}$. Then $\left\|\mathbf{z}_u\right\|^2 = \left\|\mathbf{x}_u\right\|^2$, $\left\|\mathbf{z}_v\right\|^2 = \left\|\mathbf{x}_v\right\|^2$, $\left\langle \mathbf{z}_u, \mathbf{z}_v \right\rangle = \left\langle \mathbf{x}_u, \mathbf{x}_v \right\rangle$. Moreover the included angle cosine of $\mathbf{z}_u$ and $\mathbf{z}_v$ is equal to the included angle cosine of $\mathbf{x}_u$ and $\mathbf{x}_v$. Both are $\cos\alpha$.

The proof of Theorem 3.2 can be found in Appendix C.

The following definition is used in my subsequent derivations.

Definition 2.1 $< \mathbf{\Phi x}_u, \mathbf{\Phi x}_v >= \mathbf{x}_{uI}^T(\mathbf{\Phi}_I^T\mathbf{\Phi}_I)\mathbf{x}_{vI} = \mathbf{x}_{uI}^T\mathbf{V\Lambda V}^T\mathbf{x}_{vI} = \mathbf{z}_u^T\mathbf{\Lambda z}_v = \sum_{i=1}^{|I|}\lambda_i z_{ui} z_{vi}$ where $z_{ui} \in \mathbf{z}_u$, $z_{vi} \in \mathbf{z}_v$.

Let $< \mathbf{\Phi x}_u, \mathbf{\Phi x}_v >= \sum_{i=1}^{k1}\lambda_i z_{ui} z_{vi} + \sum_{i=1}^{k2}\lambda_i z_{ui} z_{vi}$ where $|I| = k1 + k2$, $\sum_{i=1}^{k1}\lambda_i z_{ui} z_{vi}$ denotes summation of terms in which signs of $z_{ui}$ and $z_{vi}$ are same. And $\sum_{i=1}^{k1}\lambda_i z_{ui} z_{vi}$ denotes summation of terms in which signs of $z_{ui}$ and $z_{vi}$ are opposite.

If signs of elements in corresponding position of $\mathbf{z}_u$ and $\mathbf{z}_v$ are same, formula（3.3）holds.

$$\lambda_{\min}\sum_{i=1}^{|I|} z_{ui} z_{vi} \leq \sum_{i=1}^{|I|}\lambda_i z_{ui} z_{vi} \leq \lambda_{\max}\sum_{i=1}^{|I|} z_{ui} z_{vi} \tag{3.3}$$

Formula（3.3）is equivalent to $\lambda_{\min} < \mathbf{x}_u, \mathbf{x}_v > \leq < \mathbf{\Phi x}_u, \mathbf{\Phi x}_v > \leq \lambda_{\max} < \mathbf{x}_u, \mathbf{x}_v >$ where $< \mathbf{\Phi x}_u, \mathbf{\Phi x}_v >= \sum_{i=1}^{|I|}\lambda_i z_{ui} z_{vi}$, $\lambda_{\min} < \mathbf{x}_u, \mathbf{x}_v >= \lambda_{\min}\sum_{i=1}^{|I|} z_{ui} z_{vi}$, and $\lambda_{\max} < \mathbf{x}_u, \mathbf{x}_v > \lambda_{\max}\sum_{i=1}^{|I|} z_{ui} z_{vi}$. Hence $\lambda_{\min}\sum_{i=1}^{|I|} z_{ui} z_{vi} \leq \sum_{i=1}^{|I|}\lambda_i z_{ui} z_{vi} \leq \lambda_{\max}\sum_{i=1}^{|I|} z_{ui} z_{vi}$, $\lambda_{\min} < \mathbf{x}_u, \mathbf{x}_v > \leq < \mathbf{\Phi x}_u, \mathbf{\Phi x}_v > \leq \lambda_{\max} < \mathbf{x}_u, \mathbf{x}_v >$.

*C. Identical Orthant Theorem*

Theorem 3.3 (Identical Orthant Theorem): Let $\mathbf{z}_u^s, \mathbf{z}_v^s$ denote vectors which consist of $z_{ui}$ and $z_{vi}$ of the same sign, $\mathbf{z}_u^o, \mathbf{z}_v^o$ denote vectors which consist of $z_{ui}$ and $z_{vi}$ of the opposite sign. Then $\dfrac{\left|\left\langle \mathbf{z}_u^o, \mathbf{z}_v^o \right\rangle\right|}{\left\langle \mathbf{z}_u, \mathbf{z}_v \right\rangle} \leq \dfrac{1}{\cos\alpha} - 1$, where $\cos\alpha$ is the included angle cosine of $\mathbf{z}_u$ and $\mathbf{z}_v$.

The proof of Theorem 3.3 can be found in Appendix D. Orthant is quadrant in 2D space and octant in 3D space.

Theorem 3.3 (Identical Orthant Theorem) shows the proportion of $\left|\left\langle \mathbf{z}_u^o, \mathbf{z}_v^o \right\rangle\right| / \left\langle \mathbf{z}_u, \mathbf{z}_v \right\rangle$. The less the proportion is; the more elements of same signs in $\mathbf{z}_u$ and $\mathbf{z}_v$ are. If $\cos\alpha = 1$, $\mathbf{z}_u$ and $\mathbf{z}_v$ are in the same orthant. Hence theorem 3.3 is

referred to as identical orthant theorem. The larger $\cos\alpha$, the less the proportion of $\frac{\sum_{i=1}^{k2}\lambda_i z_{ui} z_{vi}}{\sum_{i=1}^{|I|}\lambda_i z_{ui} z_{vi}}$. As shown in Fig. 1(b), $\cos\alpha$ is close to 1 in the natural image. Hence $\lambda_{\min}\sum_{i=1}^{|I|} z_{ui} z_{vi} \le \sum_{i=1}^{|I|}\lambda_i z_{ui} z_{vi} \le \lambda_{\max}\sum_{i=1}^{|I|} z_{ui} z_{vi}$ and $\lambda_{\min} < \mathbf{x}_u, \mathbf{x}_v > \le < \mathbf{\Phi x}_u, \mathbf{\Phi x}_v > \le \lambda_{\max} < \mathbf{x}_u, \mathbf{x}_v >$ is easy to hold in RCP_IP.

Although theorem 3.3 (identical orthant theorem) only shows the proportion of $\frac{\left|\left\langle \mathbf{z}_u^o, \mathbf{z}_v^o \right\rangle\right|}{\left\langle \mathbf{z}_u, \mathbf{z}_v \right\rangle}$, it doesn't show number of minus terms of $\sum_{i=1}^{|I|}\lambda_i z_{ui} z_{vi}$. $\mathbf{z}_u$, $\mathbf{z}_v$, $\mathbf{x}_u$ and $\mathbf{x}_v$ are unknown, but the measurement matrix is known in CS. Hence Theorem 3.4 (Minus Term Theorem) focus from $\mathbf{z}_u$ and $\mathbf{z}_v$ to eigenvalues of $\mathbf{G}$.

### *D. Minus Term Theorem*

Theorem 3.4 (Minus Term Theorem) : $\lambda_{\min}\sum_{i=1}^{|I|} z_{ui} z_{vi} \le \sum_{i=1}^{|I|}\lambda_i z_{ui} z_{vi} \le \lambda_{\max}\sum_{i=1}^{|I|} z_{ui} z_{vi}$ holds, if $\sum_{i=1}^{k1}(\lambda_{\max}\cos\alpha - \lambda_i) z_{ui} z_{vi} \ge 0$ and $\sum_{i=1}^{k1}\left(\frac{(\lambda_{\max}-\lambda_{\min})(1-\cos\alpha)}{\cos^2\alpha} - (\lambda_i - \lambda_{\min})\right) z_{ui} z_{vi} \le 0$.

The proof of Theorem 3.4 can be found in Appendix E.

Since eigenvalues of any $\mathbf{G}$ can be calculated easily, the number of minus term can be estimated simply in $\sum_{i=1}^{k1}(\lambda_{\max}\cos\alpha - \lambda_i) z_{ui} z_{vi}$ and $\sum_{i=1}^{k1}\left(\frac{(\lambda_{\max}-\lambda_{\min})(1-\cos\alpha)}{\cos^2\alpha} - \left|\lambda_i - \lambda_{\min}\right|\right) z_{ui} z_{vi}$. Hence theorem 3.4 is referred to as minus term theorem. The less the number of minus term is, the easier $\lambda_{\min}\sum_{i=1}^{|I|} z_{ui} z_{vi} \le \sum_{i=1}^{|I|}\lambda_i z_{ui} z_{vi} \le \lambda_{\max}\sum_{i=1}^{|I|} z_{ui} z_{vi}$ holds. Theorem 3.4 (Minus Term Theorem) is important supplements of Theorem 3.3 (Identical Orthant Theorem).

## IV. RCP OF COMPRESSIBLE SIGNAL AND APPLICATION OF RCP

RIP applies to not only sparse signals but also compressible signals. If signals are sparse, RIP is RIP of the measurement matrix. Meanwhile the measurement matrix is a reconstruction matrix, too. If signals are compressible, RIP is RIP of the reconstruction matrix ($\mathbf{\Phi\Psi}^T$), where $\mathbf{\Psi}$ is a sparsity basis. In the same way, RCP applies to not only sparse signals but also compressible signals.

If the sparsity basis ($\mathbf{\Psi}$) is an orthogonal matrix (e.g. Discrete Fourier Transform (DFT), Discrete Cosine Transform (DCT) et al), energies of compressible signals are same with energies of their transform domain coefficient vectors and the included angle cosine of transform domain coefficient vectors is equal to the included angle cosine of compressible signal vectors. The proof can be found in Appendix F.

Measurements are collected by a 0-1 random matrix and the sparsity basis ($\mathbf{\Psi}$) is a DCT matrix in fig. 3 ($\mu$ is the included angle cosine of 2 adjacent columns. $\boldsymbol{\alpha}_j$ is the vector of transform domain coefficient of $\mathbf{x}_j$ where $\boldsymbol{\alpha}_j = \mathbf{\Psi x}_j$ .). As shown in fig. 3(b), the curves of $\boldsymbol{\alpha}_j$ and $\mathbf{x}_j$ are same completely. And the curve of $\mathbf{y}_j$ is similar with $\boldsymbol{\alpha}_j$ and $\mathbf{x}_j$.

Push broom imaging by a linear array of detectors is main data collection technology in remote sensing. RCP is very good prior information. New reconstruction algorithm can be designed by use of RCP. I have designed 2DOMP [6] for sparse signal and 2DDOMP[8] for compressible signal based on RCP. And they work well.

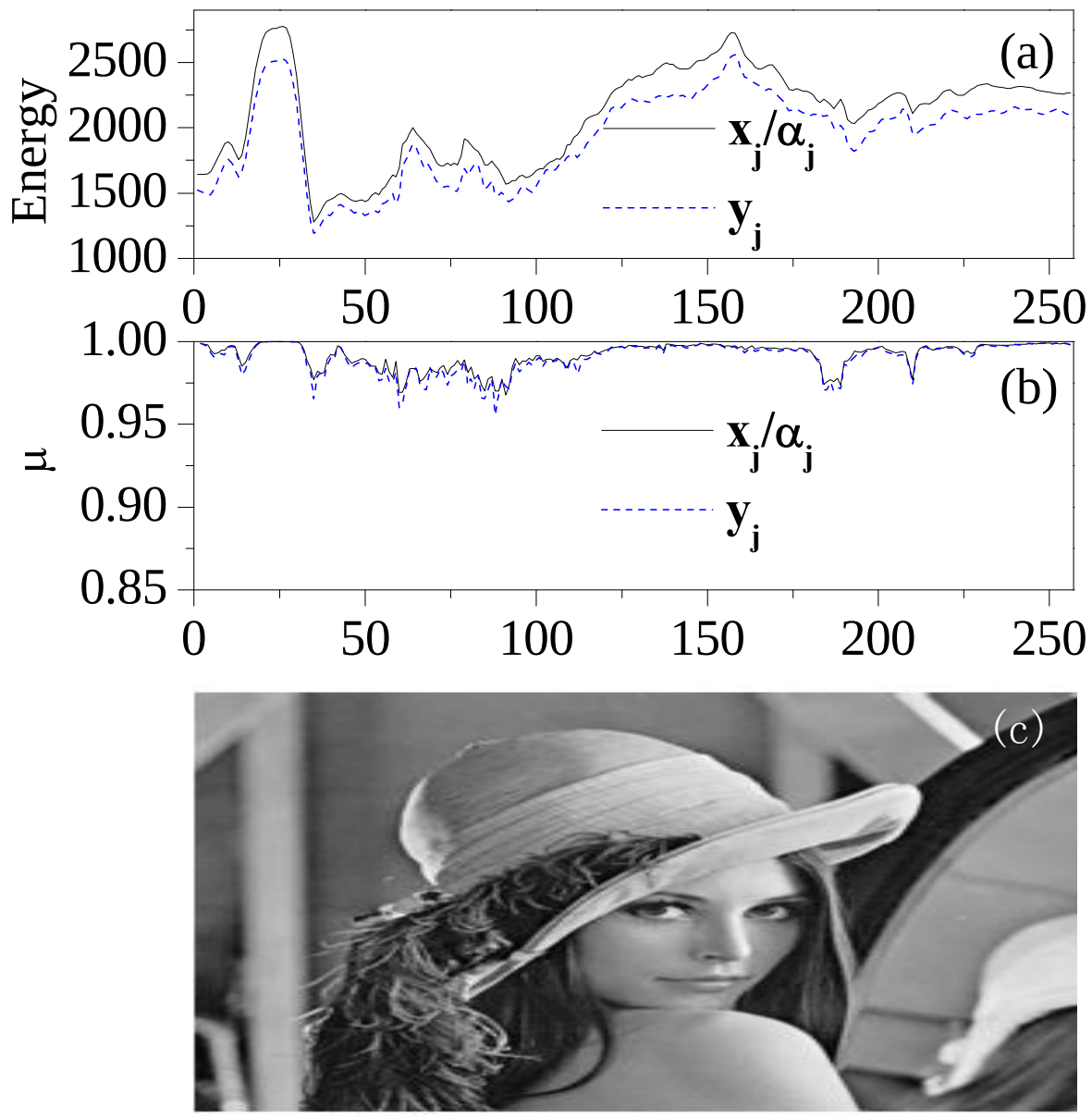

Fig. 3. Analysis curves of **X** and **Y** and gray image. (a) Energy; (b) Correlation coefficient between 2 columns; (c) Lena.

## V. DISCUSSION

### *A. Difficulty of Probability Analysis of Minus Term Theorem*

Both Theorem 3.3 (Identical Orthant Theorem) and Theorem 3.4 (Minus Term Theorem) are qualitative conclusion. Whether $\lambda_{\min}\sum_{i=1}^{|I|} z_{ui}z_{vi} \le \sum_{i=1}^{|I|}\lambda_i z_{ui}z_{vi} \le \lambda_{\max}\sum_{i=1}^{|I|} z_{ui}z_{vi}$ holds decides on the relation of $\mathbf{z}_u$, $\mathbf{z}_v$ and $\cos\alpha$ in Theorem 3.3 (Identical Orthant Theorem) and the relation of $\lambda_i$, $\cos\alpha$ and $z_{ui}z_{vi}$ in Theorem 3.4 (Minus Term Theorem). They can't show what the probability is exactly.

Since $\mathbf{z}_u$ and $\mathbf{z}_v$ is coherent in natural images and their correlation coefficient is $\cos\alpha$, the probability distribution function of $\mathbf{z}_u^T\mathbf{z}_v$ is very complicated and long even if $\mathbf{z}_u$ and $\mathbf{z}_v$ follow Gauss distribution[9]. Because natural images are miscellaneous, maybe to find a probability distribution function for $z_{ui}z_{vi}$ is difficult or impossible.

The Gram matrix $\mathbf{G}$ of a Gaussian matrix is a Wishart matrix, too. The probability density function of the joint distribution of eigenvalues of a Wishart matrix can be gained. And the function is very complicated and long, too[10]. But there is not an exact probability distribution of single eigenvalue of a Wishart matrix. Maybe a succinct and exact probability function that $\lambda_{\min}\sum_{i=1}^{|I|} z_{ui}z_{vi} \le \sum_{i=1}^{|I|}\lambda_i z_{ui}z_{vi} \le \lambda_{\max}\sum_{i=1}^{|I|} z_{ui}z_{vi}$ holds is difficult or impossible.

### *B. Approximate Statistics Analysis of Eigenvalues of Wishart Matrix*

Gaussian matrices are the most studied in CS, so the scope of study is limited in Gaussian matrices here. Although the exact probability analysis of eigenvalues of a Wishart matrix is difficult, the approximate probability distribution estimation is easy by statistical analysis. Eigenvalues of $\mathbf{G}$ follow Wishart distribution. And diagonal elements of $\mathbf{G}$ follow chi-squared distribution. Let an i.i.d. Gaussian matrix $\mathbf{\Phi} \in \mathbf{R}^{M\times N}$ ($M < N$), $\mathbf{\Phi}_{ij} \sim N(0,\frac{1}{M})$, $i \in [1, M]$, $j \in [1, N]$. Because of $MG_{ii} \sim \chi^2(M)$, the mathematical expectation and the variance of $MG_{ii}$ are $E(MG_{ii}) = M$ and $D(MG_{ii}) = 2M$. $E(G_{ii}) = 1$, since $\sum_{i=1}^{|I|}\lambda_i = tr(\mathbf{\Phi}_I^T\mathbf{\Phi}_I) = \sum_{i=1}^{|I|} G_{ii}$.

If normalizing all columns of $\mathbf{\Phi}$, based on Gershgorin circle theorem[11], all eigenvalues of $\mathbf{G}$ are in Gershgorin circle such that $\left|\lambda_i - 1\right| \le \max \sum_{i \ne j} \left|\mathbf{G}_{ij}\right|$ holds, where $j = 1, 2, \cdots, |I|; i = 1, 2, \cdots, |I|$. Radius of Gershgorin circle can be shorted by optimizing matrices [2, 3], too. The better independence of all columns, the shorter the radius of Gershgorin circle such that all eigenvalues are close to 1. Hence eigenvalues of $\mathbf{G}$ should follow the Gauss distribution approximately based on the law of large numbers.

I found that the mean and the variance of $\sqrt{\lambda_i}$ are close to 1 and $\frac{|I|}{2M}$ respectively by lots of statistical analysis. Moreover $(\sqrt{\lambda_i} - 1)\sqrt{\frac{2M}{|I|}}$ follow Gauss distribution by Jarque–Bera test and $(\sqrt{\lambda_i} - 1)\sqrt{\frac{2M}{|I|}} \sim N(0,1)$ by Kolmogorov–Smirnov test if M=128 and N=256. More results of numerical experiments are presented for N=256, N=512, N=1024 and N=2048 in fig. 4. The zones below the curves have passed 1000 Kolmogorov– Smirnov tests, while the zones above the curves are on the contrary in fig. 4.

$|I|$ and M are very little in order to gain higher compression ratio and good reconstruction effect in most CS. N is little in order to reduce the size of matrices and to improve the computing speed, too. Therefore, the zones below the curves pass 1000 Kolmogorov–Smirnov tests are large in fig. 4. The assumption on $(\sqrt{\lambda_i} - 1)\sqrt{\frac{2M}{K}} \sim N(0,1)$ is feasible empirically for little matrices with little N. Then $P(\lambda_i \ge \lambda_{\max}\cos\alpha) = 1 - \frac{1}{\sqrt{2\pi}} \int_{-\infty}^{(\sqrt{\lambda_{\max}\cos\alpha} - 1)\sqrt{\frac{2M}{|I|}}} e^{-\frac{t^2}{2}} dt$ and $P(\lambda_i \le [\lambda_{\min} + (\lambda_{\max} - \lambda_{\min})\frac{1-\cos\alpha}{\cos^2\alpha}]) = \frac{1}{\sqrt{2\pi}} \int_{-\infty}^{(\sqrt{[\lambda_{\min} + (\lambda_{\max} - \lambda_{\min})\frac{1-\cos\alpha}{\cos^2\alpha}]} - 1)\sqrt{M}} e^{-\frac{t^2}{2}} dt$. If $\cos\alpha$ is close to 1, $\sum_{i=1}^{k1} (\lambda_{\max}\cos\alpha - \lambda_i) z_{ui} z_{vi} \ge 0$ and $\sum_{i=1}^{k1} (\frac{(\lambda_{\max} - \lambda_{\min})(1 - \cos\alpha)}{\cos^2\alpha} - (\lambda_i - \lambda_{\min}))\, z_{ui} z_{vi} \le 0$ are easy to hold in Theorem 3.4 (Minus Term Theorem).

In real application case of CS, reconstruction matrices are known, so their eigenvalues can be gained beforehand. Thus exact probability distribution of their eigenvalues is known. Therefore good matrices can be gained by screening lots of matrices.

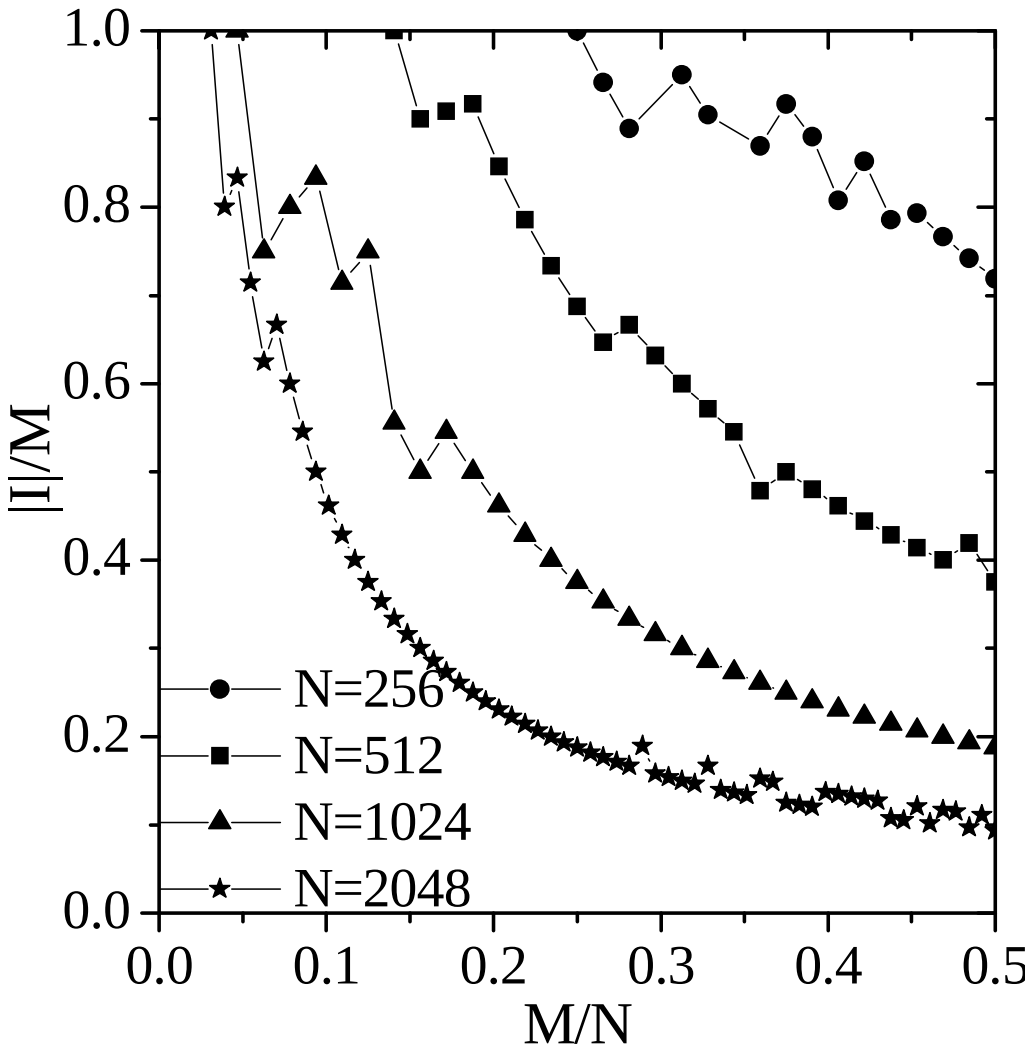

Fig. 4. Analysis curves of Kolmogorov–Smirnov test for different N.

## VI. CONCLUSION

Measurements can retain approximately not only energies but also directions of signals in compressive sensing. RCP (RCP_JL and RCP_IP) is proposed and demonstrated to depict the inheritance relationships of directions of signals in this paper. RCP is important supplement and development of RIP.

Natural images have good prior information such as spatio-temporal continuity and structure. Energies and directions of natural images' adjacent columns are almost same, and their changes are slow. Hence measurement and reconstruction matrices show good RCP when natural images are collected based on linear array push-broom mode. RCP provides new theory for designing new reconstruction algorithms.

## APPENDIX

### A. Proof of Theorem 2.1（RCP_JL）

Since $\mathbf{y}_u = \mathbf{\Phi x}_u$ and $\mathbf{y}_v = \mathbf{\Phi x}_v$, by Johnson-Lindenstrauss lemma,

$$(1-\varepsilon)\|\mathbf{x}_u - \mathbf{x}_v\|^2 \le \|\mathbf{y}_u - \mathbf{y}_v\|^2 \le (1+\varepsilon)\|\mathbf{x}_u - \mathbf{x}_v\|^2 .$$

By the law of cosines,

$$\|\mathbf{y}_u - \mathbf{y}_v\|^2 = \|\mathbf{y}_u\|^2 + \|\mathbf{y}_v\|^2 - 2\|\mathbf{y}_u\|\|\mathbf{y}_v\|\cos\beta$$

$$\|\mathbf{x}_u - \mathbf{x}_v\|^2 = \|\mathbf{x}_u\|^2 + \|\mathbf{x}_v\|^2 - 2\|\mathbf{x}_u\|\|\mathbf{x}_v\|\cos\alpha .$$

Therefore

$$\frac{(1-\varepsilon)\|\mathbf{x}_u\|^2 + (1-\varepsilon)\|\mathbf{x}_v\|^2 - 2(1-\varepsilon)\|\mathbf{x}_u\|\|\mathbf{x}_v\|\cos\alpha - (\|\mathbf{y}_u\|^2 + \|\mathbf{y}_v\|^2)}{2\|\mathbf{y}_u\|\|\mathbf{y}_v\|}$$

$$\ge \cos\beta \ge \frac{(1+\varepsilon)\|\mathbf{x}_u\|^2 + (1+\varepsilon)\|\mathbf{x}_v\|^2 - 2(1+\varepsilon)\|\mathbf{x}_u\|\|\mathbf{x}_v\|\cos\alpha - (\|\mathbf{y}_u\|^2 + \|\mathbf{y}_v\|^2)}{2\|\mathbf{y}_u\|\|\mathbf{y}_v\|}$$

Note that the RIP implies that

$$\frac{(-\varepsilon+\delta_{Ku})\|\mathbf{x}_u\|^2 + (-\varepsilon+\delta_{Kv})\|\mathbf{x}_v\|^2 - 2(1-\varepsilon)\|\mathbf{x}_u\|\|\mathbf{x}_v\|\cos\alpha}{2\sqrt{(1-\delta_{Ku})(1-\delta_{Kv})}\,\|\mathbf{x}_u\|\|\mathbf{x}_v\|}$$

$$\ge \cos\beta \ge \frac{(\varepsilon-\delta_{Ku})\|\mathbf{x}_u\|^2 + (\varepsilon-\delta_{Kv})\|\mathbf{x}_v\|^2 - 2(1+\varepsilon)\|\mathbf{x}_u\|\|\mathbf{x}_v\|\cos\alpha}{2\sqrt{(1+\delta_{Ku})(1+\delta_{Kv})}\,\|\mathbf{x}_u\|\|\mathbf{x}_v\|}$$

Since $\delta_{\max} = \max(\delta_{Ku}, \delta_{Kv})$,

$$\frac{(\delta_{\max}-\varepsilon)(\|\mathbf{x}_u\|^2 + \|\mathbf{x}_v\|^2)}{2(1-\delta_{\max})\|\mathbf{x}_u\|\|\mathbf{x}_v\|} + \frac{1-\varepsilon}{1-\delta_{\max}}\cos\alpha \ge \cos\beta \ge \frac{(\varepsilon-\delta_{\max})(\|\mathbf{x}_u\|^2 + \|\mathbf{x}_v\|^2)}{2(1+\delta_{\max})\|\mathbf{x}_u\|\|\mathbf{x}_v\|} + \frac{1+\varepsilon}{1+\delta_{\max}}\cos\alpha .$$

Since $\xi = \dfrac{\|\mathbf{x}_u\|^2 + \|\mathbf{x}_v\|^2}{2\|\mathbf{x}_u\|\|\mathbf{x}_v\|}$,

$$\frac{\delta_{\max}-\varepsilon}{1-\delta_{\max}}\xi + \frac{1-\varepsilon}{1-\delta_{\max}}\cos\alpha \ge \cos\beta \ge \frac{\varepsilon-\delta_{\max}}{1+\delta_{\max}}\xi + \frac{1+\varepsilon}{1+\delta_{\max}}\cos\alpha ,$$

which completes the proof.

### B. Proof of Theorem 3.1 (RCP_IP)

#### 1). Proof of Theorem 3.1(1)

Under the assumption of $\lambda_{\min} < \mathbf{x}_u, \mathbf{x}_v > \le < \mathbf{\Phi x}_u, \mathbf{\Phi x}_v > \le \lambda_{\max} < \mathbf{x}_u, \mathbf{x}_v >$, put $\mathbf{y}_u = \mathbf{\Phi x}_u$ and $\mathbf{y}_v = \mathbf{\Phi x}_v$, then

$$\lambda_{\min}\|\mathbf{x}_u\|\cdot\|\mathbf{x}_v\|\cos\alpha \le \|\mathbf{y}_u\|\cdot\|\mathbf{y}_v\|\cos\beta \le \lambda_{\max}\|\mathbf{x}_u\|\cdot\|\mathbf{x}_v\|\cos\alpha$$

$$\lambda_{\min}\frac{\|\mathbf{x}_u\|\cdot\|\mathbf{x}_v\|}{\|\mathbf{y}_u\|\cdot\|\mathbf{y}_v\|}\cos\alpha \le \cos\beta \le \lambda_{\max}\frac{\|\mathbf{x}_u\|\cdot\|\mathbf{x}_v\|}{\|\mathbf{y}_u\|\cdot\|\mathbf{y}_v\|}\cos\alpha .$$

Note that the RIP implies that

$$\frac{\lambda_{\min}}{\sqrt{(1+\delta_{Ku})\,(1+\delta_{Kv})}}\cos\alpha \le \cos\beta \le \frac{\lambda_{\max}}{\sqrt{(1-\delta_{Ku})\,(1-\delta_{Kv})}}\cos\alpha .$$

Suppose $\lambda_{\min} \geq 1-\delta_K$ , $\lambda_{\max} \leq 1+\delta_K$ , $\delta_K \in$（0, 1） based on RIP, then

$$\frac{1-\delta_K}{\sqrt{(1+\delta_{Ku})(1+\delta_{Kv})}}\cos\alpha \leq \cos\beta \leq \frac{1+\delta_K}{\sqrt{(1-\delta_{Ku})(1-\delta_{Kv})}}\cos\alpha$$

$$\frac{1-\delta_K}{1+\delta_K}\cos\alpha \leq \cos\beta \leq \frac{1+\delta_K}{1-\delta_K}\cos\alpha ,$$

which completes the proof.

2). *Proof of Theorem 3.1(2)*

Note that the RIP implies that

$$(1-\delta_K)\|\mathbf{x}_u+\mathbf{x}_v\|^2 \leq \|\mathbf{y}_u+\mathbf{y}_v\|^2 \leq (1+\delta_K)\|\mathbf{x}_u+\mathbf{x}_v\|^2 ,$$

and $(1-\delta_K)\|\mathbf{x}_u-\mathbf{x}_v\|^2 \leq \|\mathbf{y}_u-\mathbf{y}_v\|^2 \leq (1+\delta_K)\|\mathbf{x}_u-\mathbf{x}_v\|^2$.

Following from the parallelogram identity,

$$\langle \mathbf{y}_u, \mathbf{y}_v \rangle = \frac{\|\mathbf{y}_u+\mathbf{y}_v\|^2-\|\mathbf{y}_u-\mathbf{y}_v\|^2}{4} .$$

Since $I_u$ and $I_v$ are disjoint sets, $I_u \cap I_v = \varnothing$ and $\langle \mathbf{x}_u, \mathbf{x}_v \rangle = 0$ ,

$$-\frac{\delta_K(\|\mathbf{x}_u\|^2+\|\mathbf{x}_v\|^2)}{2} \leq \langle \mathbf{y}_u, \mathbf{y}_v \rangle \leq \frac{\delta_K(\|\mathbf{x}_u\|^2+\|\mathbf{x}_v\|^2)}{2}$$

$$-\delta_K(\|\mathbf{x}_u\|\cdot\|\mathbf{x}_v\| \leq \langle \mathbf{y}_u, \mathbf{y}_v \rangle \leq \delta_K(\|\mathbf{x}_u\|\cdot\|\mathbf{x}_v\| .$$

Since $(1-\delta_{Ku})\|\mathbf{x}_u\|^2 \leq \|\mathbf{y}_u\|^2 \leq (1+\delta_{Ku})\|\mathbf{x}_u\|^2$ and $(1-\delta_{Kv})\|\mathbf{x}_v\|^2 \leq \|\mathbf{y}_v\|^2 \leq (1+\delta_{Kv})\|\mathbf{x}_v\|^2$ ,

$$\frac{-\delta_K}{\sqrt{1+\delta_{Ku}}\sqrt{1+\delta_{Kv}}} \leq \cos\beta \leq \frac{\delta_K}{\sqrt{1-\delta_{Ku}}\sqrt{1-\delta_{Kv}}}$$

$$\frac{-\delta_K}{1+\delta_{\max}} \leq \cos\beta \leq \frac{\delta_K}{1-\delta_{\max}} ,$$

which completes the proof.

C. *Proof of Theorem 3.2*

Since $\mathbf{V}$ is an orthogonal matrix,

$$\|\mathbf{z}_u\|^2 = (\mathbf{V}^T\mathbf{x}_{uI})^T\mathbf{V}^T\mathbf{x}_{uI} = \mathbf{x}_{uI}{}^T(\mathbf{V}\mathbf{V}^T)\mathbf{x}_{uI} = \|\mathbf{x}_{uI}\|^2 = \|\mathbf{x}_u\|^2 .$$

Similarly,

$$\|\mathbf{z}_v\|^2 = \|\mathbf{x}_v\|^2 ,\ \langle \mathbf{z}_u, \mathbf{z}_v \rangle = (\mathbf{V}^T\mathbf{x}_{uI})^T\mathbf{V}^T\mathbf{x}_{vI} = \mathbf{x}_{uI}{}^T(\mathbf{V}\mathbf{V}^T)\mathbf{x}_{vI} = \langle \mathbf{x}_{uI}, \mathbf{x}_{vI} \rangle = \langle \mathbf{x}_u, \mathbf{x}_v \rangle ,$$

and $\dfrac{\langle \mathbf{z}_u, \mathbf{z}_v \rangle}{\|\mathbf{z}_u\|\cdot\|\mathbf{z}_v\|} = \dfrac{\langle \mathbf{x}_u, \mathbf{x}_v \rangle}{\|\mathbf{x}_u\|\cdot\|\mathbf{x}_v\|} = \cos\alpha$ ,

which completes the proof.

D. *Proof of Theorem 3.3 (Identical Orthant Theorem)*

Let $\cos\theta$ be the included angle cosine of the vector pair ($\mathbf{z}_u^s, \mathbf{z}_v^s$), $\cos\gamma$ be the included angle cosine of the vector pair ($\mathbf{z}_u^o, \mathbf{z}_v^o$).

1). $\because \langle \mathbf{z}_u^s, \mathbf{z}_v^s \rangle + \langle \mathbf{z}_v^o, \mathbf{z}_v^o \rangle = \langle \mathbf{z}_u, \mathbf{z}_v \rangle$ where $\langle \mathbf{z}_u^o, \mathbf{z}_v^o \rangle \leq 0$

$\therefore \langle \mathbf{z}_u^s, \mathbf{z}_v^s \rangle \geq \langle \mathbf{z}_u, \mathbf{z}_v \rangle$

$\therefore \left\|\mathbf{z}_u^s\right\| \cdot \left\|\mathbf{z}_v^s\right\| \cos\theta \ge \left\|\mathbf{z}_u\right\| \cdot \left\|\mathbf{z}_v\right\| \cos\alpha$, where $\left\|\mathbf{z}_u^s\right\| \le \left\|\mathbf{z}_u\right\|$, $\left\|\mathbf{z}_v^s\right\| \le \left\|\mathbf{z}_v\right\|$

$\therefore \cos\theta \ge \cos\alpha$

$\because \left\langle \mathbf{z}_u, \mathbf{z}_v \right\rangle \ge \left\langle \mathbf{z}_u^o, \mathbf{z}_v^o \right\rangle$

$\because \left\|\mathbf{z}_u\right\| \cdot \left\|\mathbf{z}_v\right\| \cos\alpha \ge \left\|\mathbf{z}_u^o\right\| \cdot \left\|\mathbf{z}_v^o\right\| \cos\gamma$, $\left\|\mathbf{z}_u^o\right\| \le \left\|\mathbf{z}_u\right\|$, $\left\|\mathbf{z}_v^o\right\| \le \left\|\mathbf{z}_v\right\|$, and $\cos\gamma \le 0$

$\therefore \cos\alpha \ge \cos\gamma$

$\therefore \cos\theta \ge \cos\alpha \ge \cos\gamma$.

2). $\because \left\|\mathbf{z}_u^s\right\| \cdot \left\|\mathbf{z}_v^s\right\| \cos\theta + \left\|\mathbf{z}_u^o\right\| \cdot \left\|\mathbf{z}_v^o\right\| \cos\gamma = \left\|\mathbf{z}_u\right\| \cdot \left\|\mathbf{z}_v\right\| \cos\alpha$

$$\therefore \frac{\left\|\mathbf{z}_u^s\right\| \cdot \left\|\mathbf{z}_v^s\right\|}{\left\|\mathbf{z}_u\right\| \cdot \left\|\mathbf{z}_v\right\|} \cos\theta - \frac{\left\|\mathbf{z}_u^o\right\| \cdot \left\|\mathbf{z}_v^o\right\|}{\left\|\mathbf{z}_u\right\| \cdot \left\|\mathbf{z}_v\right\|} \left|\cos\gamma\right| = \cos\alpha$$

$$\therefore 1 \ge \frac{\left\|\mathbf{z}_u^s\right\| \cdot \left\|\mathbf{z}_v^s\right\|}{\left\|\mathbf{z}_u\right\| \cdot \left\|\mathbf{z}_v\right\|} \ge \cos\alpha .$$

3). $$\because \frac{\left\langle \mathbf{z}_u^s, \mathbf{z}_v^s \right\rangle}{\left\langle \mathbf{z}_u, \mathbf{z}_v \right\rangle} = \frac{\left\|\mathbf{z}_u^s\right\| \cdot \left\|\mathbf{z}_v^s\right\| \cos\theta}{\left\|\mathbf{z}_u\right\| \cdot \left\|\mathbf{z}_v\right\| \cos\alpha} \quad \text{and} \quad 1 \ge \frac{\left\|\mathbf{z}_u^s\right\| \cdot \left\|\mathbf{z}_v^s\right\|}{\left\|\mathbf{z}_u\right\| \cdot \left\|\mathbf{z}_v\right\|} \ge \cos\alpha$$

$$\therefore \frac{\cos\alpha}{\cos\gamma} \ge \frac{\left\langle \mathbf{z}_u^s, \mathbf{z}_v^s \right\rangle}{\left\langle \mathbf{z}_u, \mathbf{z}_v \right\rangle} \ge \cos\theta \quad \text{and} \quad \cos\theta \ge \cos\alpha \ge \cos\gamma$$

$$\therefore \frac{1}{\cos\alpha} \ge \frac{\left\langle \mathbf{z}_u^s, \mathbf{z}_v^s \right\rangle}{\left\langle \mathbf{z}_u, \mathbf{z}_v \right\rangle} \ge \cos\alpha .$$

4). $$\because \frac{\left\langle \mathbf{z}_u^o, \mathbf{z}_v^o \right\rangle}{\left\langle \mathbf{z}_u, \mathbf{z}_v \right\rangle} = \frac{\left\langle \mathbf{z}_u, \mathbf{z}_v \right\rangle - \left\langle \mathbf{z}_u^s, \mathbf{z}_v^s \right\rangle}{\left\langle \mathbf{z}_u, \mathbf{z}_v \right\rangle} = 1 - \frac{\left\langle \mathbf{z}_u^s, \mathbf{z}_v^s \right\rangle}{\left\langle \mathbf{z}_u, \mathbf{z}_v \right\rangle} \quad \text{and} \quad \frac{1}{\cos\alpha} \ge \frac{\left\langle \mathbf{z}_u^s, \mathbf{z}_v^s \right\rangle}{\left\langle \mathbf{z}_u, \mathbf{z}_v \right\rangle} \ge \cos\alpha$$

$$\therefore \frac{\left\langle \mathbf{z}_u^o, \mathbf{z}_v^o \right\rangle}{\left\langle \mathbf{z}_u, \mathbf{z}_v \right\rangle} \ge 1 - \frac{1}{\cos\alpha} .$$

$$\because 0 \ge \frac{\left\langle \mathbf{z}_1^y, \mathbf{z}_2^y \right\rangle}{\left\langle \mathbf{z}_1, \mathbf{z}_2 \right\rangle},$$

$$\therefore \frac{\left|\left\langle \mathbf{z}_u^o, \mathbf{z}_v^o \right\rangle\right|}{\left\langle \mathbf{z}_u, \mathbf{z}_v \right\rangle} \le \frac{1}{\cos\alpha} - 1,$$

which completes the proof.

*E. Proof of Theorem 3.4 (Minus Term Theorem)*

1) By Theorem 3.3, $\frac{1}{\cos\alpha} \ge \frac{\left\langle \mathbf{z}_u^s, \mathbf{z}_v^s \right\rangle}{\left\langle \mathbf{z}_u, \mathbf{z}_v \right\rangle} \ge \cos\alpha$,

and therefore

$$\lambda_{\max} \sum_{i=1}^{|I|} z_{ui} z_{vi} - \sum_{i=1}^{|I|} \lambda_i z_{ui} z_{vi}$$

$$\geq \lambda_{\max} \cos\alpha \sum_{i=1}^{k1} z_{ui} z_{vi} - \sum_{i=1}^{k1} \lambda_i z_{ui} z_{vi}$$

$$= \sum_{i=1}^{k1} (\lambda_{\max} \cos\alpha - \lambda_i) z_{ui} z_{vi} .$$

2）By Theorem 3.3, $\frac{\left|\left\langle \mathbf{z}_u^o, \mathbf{z}_v^o \right\rangle\right|}{\left\langle \mathbf{z}_u, \mathbf{z}_v \right\rangle} \leq \frac{1}{\cos\alpha} - 1$ and $\frac{1}{\cos\alpha} \geq \frac{\left\langle \mathbf{z}_u^s, \mathbf{z}_v^s \right\rangle}{\left\langle \mathbf{z}_u, \mathbf{z}_v \right\rangle} \geq \cos\alpha$,

and therefore

$$\lambda_{\min} \sum_{i=1}^{|I|} z_{ui} z_{vi} - \sum_{i=1}^{|I|} \lambda_i z_{ui} z_{vi}$$

$$\leq \lambda_{\min} \sum_{i=1}^{k1} z_{ui} z_{vi} - \sum_{i=1}^{k1} \lambda_i z_{ui} z_{vi} + \sum_{i=1}^{k2} (\lambda_i - \lambda_{\min}) \left| z_{ui} z_{vi} \right|$$

$$\leq \lambda_{\min} \sum_{i=1}^{k1} z_{ui} z_{vi} - \sum_{i=1}^{k1} \lambda_i z_{ui} z_{vi} + (\lambda_{\max} - \lambda_{\min}) \frac{1-\cos\alpha}{\cos^2\alpha} \sum_{i=1}^{k1} z_{ui} z_{vi}$$

$$= \sum_{i=1}^{k1} \left( \frac{(\lambda_{\max} - \lambda_{\min})(1-\cos\alpha)}{\cos^2\alpha} - (\lambda_i - \lambda_{\min}) \right) z_{ui} z_{vi} .$$

*F. Proof:*

Since $\mathbf{\Psi} \in R^{N\times N}$ is an orthogonal matrix (i.e. an orthonormal sparsity basis, e.g. DCT and DFT matrices et al), $\mathbf{x}_u = \mathbf{\Psi}^T \boldsymbol{\alpha}_u$, and $\mathbf{x}_v = \mathbf{\Psi}^T \boldsymbol{\alpha}_v$,

$$\left\|\mathbf{x}_u\right\|^2 = \mathbf{x}_u^T \mathbf{x}_u = (\mathbf{\Psi}^T \boldsymbol{\alpha}_u)^T \mathbf{\Psi}^T \boldsymbol{\alpha}_u = \boldsymbol{\alpha}_u^T (\mathbf{\Psi}\mathbf{\Psi}^T) \boldsymbol{\alpha}_u = \boldsymbol{\alpha}_u^T \boldsymbol{\alpha}_u = \left\|\boldsymbol{\alpha}_u\right\|^2 .$$

Similarly,

$$\left\|\mathbf{x}_v\right\|^2 = \left\|\boldsymbol{\alpha}_v\right\|^2, \ \left\langle \mathbf{x}_u, \mathbf{x}_v \right\rangle = \mathbf{x}_v^T \mathbf{x}_u = (\mathbf{\Psi}^T \boldsymbol{\alpha}_v)^T \mathbf{\Psi}^T \boldsymbol{\alpha}_u = \boldsymbol{\alpha}_v^T (\mathbf{\Psi}\mathbf{\Psi}^T) \boldsymbol{\alpha}_u = \boldsymbol{\alpha}_v^T \boldsymbol{\alpha}_u = \left\langle \boldsymbol{\alpha}_u, \boldsymbol{\alpha}_v \right\rangle$$

and $\frac{\left\langle \boldsymbol{\alpha}_u, \boldsymbol{\alpha}_v \right\rangle}{\left\|\boldsymbol{\alpha}_u\right\| \cdot \left\|\boldsymbol{\alpha}_v\right\|} = \frac{\left\langle \mathbf{x}_u, \mathbf{x}_v \right\rangle}{\left\|\mathbf{x}_u\right\| \cdot \left\|\mathbf{x}_v\right\|} = \cos\alpha$

which completes the proof.